\documentclass[a4paper,11pt]{article}
\pdfoutput=1

\usepackage{amsmath}
\usepackage{graphicx}
\usepackage{jheppub}
\usepackage{xspace}
\usepackage{booktabs}
\usepackage{wasysym}
\newcommand{\MS}{\ensuremath{\overline{\text{MS}}}\xspace}

\makeatletter
\let\Re\@undefined
\makeatother
\DeclareMathOperator{\Re}{Re}

\hypersetup{
  pdftitle={
Relations between electroweak parameters
to order $\alpha_s^3$},
}

\graphicspath{{figures/}{build/figures/}}

\title{%
  \boldmath
Relations between electroweak parameters
to order~$\alpha_s^3$
}

\author[a]{Andreas Maier}
\emailAdd{andreas.maier@ifj.edu.pl}
\author[b]{and Peter Marquard}
\emailAdd{peter.marquard@desy.de}
\affiliation[a]{%
The Henryk Niewodniczański Institute of Nuclear Physics,
ul. Radzikowskiego 152, 31-342 Kraków, Poland
}
\affiliation[b]{%
  Deutsches Elektronen-Synchrotron DESY, Platanenallee 6, 15738 Zeuthen, Germany
}

\abstract{%
  Relations between electroweak parameters can be measured with
  ever-increasing precision and a future lepton collider will require
  major advances in the theoretical predictions. We derive top-quark
  mediated four-loop QCD corrections to $\Delta r$ and $\Delta \kappa$,
  that is to the relations between the weak boson masses, the
  electroweak couplings, and the weak mixing angle. To this end, we
  expand the vacuum polarization functions of charged and neutral gauge
  bosons in $M_Z^2 \ll m_t^2$, finding a negligible truncation error
  after six terms. Our results imply a shift of a few hundred keV in the
  W boson mass, which will be numerically relevant for measurements at
  the FCC-ee.
}

\preprint{DESY-26-122}

\begin{document}
\maketitle

\section{Introduction}
\label{sec:intro}

Electroweak precision observables have been instrumental in
establishing the Standard Model of particle physics. They have been
providing powerful constraints in unexplored areas, for instance
guiding Higgs searches by predicting the allowed mass
range~\cite{ALEPH:2010aa} prior to
discovery~\cite{ATLAS:2012yve,CMS:2012qbp}, and setting bounds on the
Higgs self-coupling~\cite{Degrassi:2017ucl,Kribs:2017znd}
complementary to Higgs boson pair production measurements. Nowadays,
their main role is arguable in constraining new physics scenarios and
in stress testing the Standard Model through precision measurements of
predicted relations between these observables.

Experimental measurements of observables like the W boson mass have
become extremely competitive at hadron
colliders~\cite{CDF:2022hxs,ATLAS:2024erm,CMS:2024lrd}, with relative
uncertainties of less than $10^{-4}$. Further improvements are
expected at the HL-LHC. Future lepton colliders, such as the FCC-ee,
promise even higher precision at the sub-MeV level~\cite{FCC:2025lpp}
from W pair production measurements. This will be complemented by
similarly precise measurements of a range of other electroweak
precision observables at the Z boson peak.

To match the projected experimental precision, corresponding advances
are needed on the theory side. In this work we provide four-loop QCD
corrections for two electroweak precision parameters, namely $\Delta
r$ and $\Delta \kappa$~\cite{Sirlin:1980nh,Marciano:1980pb}. $\Delta
r$ parametrizes the radiative corrections to the relation between the
W and Z boson masses $M_W, M_Z$, the Fermi constant $G_F$, and the
finestructure constant $\alpha$:
\begin{equation}
  \label{eq:deltar_def}
  (1+\Delta r) \pi \alpha = \sqrt{2} G_F M_W^2\left(1 - \frac{M_W^2}{M_Z^2}\right)
\end{equation}
$\Delta \kappa$ relates the weak mixing angle $\theta_w$ to the ratio
of the weak boson masses, viz.
\begin{equation}
  \label{eq:deltakappa_def}
  \sin^2 \theta_w = (1 + \Delta \kappa) \left(1 - \frac{M_W^2}{M_Z^2}\right).
\end{equation}
For both parameters, electroweak corrections are known at two-loop
order~\cite{Freitas:2000gg,Awramik:2002wn,Awramik:2002vu,Onishchenko:2002ve,Awramik:2003ee,Dubovyk:2018rlg}
with partial three-loop results including also mixed QCD-electroweak
corrections~\cite{Chen:2020xzx,Chen:2020xot,Dubovyk:2026nhx}. Focusing
on the pure QCD corrections, $\Delta r$ and $\Delta \kappa$ can be
expressed in terms of the transverse vector boson polarization functions
\begin{align}
  \label{eq:deltar_pi}
  \Delta r ={}& \frac{c^2}{s^2}\Re
     \left(\frac{\Pi_{WW}(M_W^2)}{M_W^2}-\frac{\Pi_{ZZ}(M_Z^2)}{M_Z^2}\right)
     + \lim_{q^2 \to 0} \frac{\Pi_{\gamma  \gamma}(q^2)}{q^2}
                + \frac{\Pi_{WW}(0)-\Re\Pi_{WW}(M_W^2)}{M_W^2},\\
  \Delta \kappa ={}& -\frac{c}{s} \frac{\Pi_{\gamma Z}(M_Z^2)}{M_Z^2} - \frac{c^2}{s^2}\Re\left(\frac{\Pi_{WW}(M_W^2)}{M_W^2}-\frac{\Pi_{ZZ}(M_Z^2)}{M_Z^2}\right),
  \label{eq:deltakappa_pi}
\end{align}
where $c = M_W/M_Z, s = \sqrt{1-c^2}$. To avoid non-perturbative
low-energy behaviour perturbative predictions focus on
\begin{equation}
  \label{eq:deltarbar}
  \Delta \bar{r} = \Delta r - \lim_{q^2 \to 0} \frac{\Pi_{\gamma  \gamma}(q^2)}{q^2} + \frac{\Re \Pi_{\gamma  \gamma}(M_Z^2)}{M_Z^2}
\end{equation}
instead of $\Delta r$. The dominant contributions arise from diagrams
involving virtual top quarks, where virtual bottom quarks are included
as well to cancel the axial anomaly. This
suggests a kinematic expansion in $M_W \sim M_Z \ll m_t$, with
the leading terms corresponding to the corrections to the parameter
$\rho = 1 + \Delta \rho$:
\begin{align}
  \Delta \rho ={}& \frac{\Pi_{ZZ}(0)}{M_Z^2}-\frac{\Pi_{WW}(0)}{M_W^2},\\
  \Delta \bar{r} ={}& -\frac{c^2}{s^2} \Delta \rho + \mathcal{O}\left(\frac{M_Z^2}{m_t^2}\right),\\
  \Delta \kappa ={}& \frac{c^2}{s^2} \Delta \rho + \mathcal{O}\left(\frac{M_Z^2}{m_t^2}\right).
\end{align}
Top quark corrections to the $\rho$ parameter are known to four loops
in
QCD~\cite{Veltman:1977kh,Djouadi:1987gn,Djouadi:1987di,Kniehl:1988ie,Avdeev:1994db,Chetyrkin:1995ix,Schroder:2005db,Chetyrkin:2006bj,Boughezal:2006xk},
and the most important three-loop electroweak and mixed
QCD-electroweak effects are available as
well~\cite{vanderBij:1986hy,Barbieri:1992nz,Barbieri:1992dq,Fleischer:1993ub,Fleischer:1994cb,vanderBij:2000cg,Faisst:2003px}. Subleading
terms in the large top-mass expansions for $\Delta \bar{r}$ and
$\Delta \kappa$ have been included at two~\cite{Halzen:1991ik} and
three loops~\cite{Chetyrkin:1995js} and have been confirmed in a
recent reanalysis~\cite{Pati:2026bmf}. The relative size of the higher
order terms computed so far suggests that their contribution at four
loops will be comparable to the projected experimental uncertainty at
the FCC-ee~\cite{FCC:2025lpp}.

The weak boson polarization functions that determine $\Delta \bar{r}$ and
$\Delta \kappa$ are conveniently expressed in terms of current
correlator polarization functions,
\begin{align}
  \label{eq:pi_weak}
  \Pi_{WW} ={}& \frac{g^2}{8} \Big[\Pi_{tb,tb}^V(q^2) + \Pi_{tb,tb}^A(q^2) \Big],
\\
\Pi_{ZZ} ={}& \frac{g^2}{16c^2}
             \sum_{q,q'=t,b}\Big[v_q v_{q'}\Pi_{qq,q'q'}^V(q^2) + \Pi_{qq,q'q'}^{A}(q^2)\Big],
\\
\Pi_{\gamma\gamma} ={}& g^2 s^2 \sum_{q,q'=t,b} Q_q Q_{q'} \Pi_{qq,q'q'}^{V}(q^2),
\label{polfun}\\
\Pi_{\gamma Z} ={}& -\frac{g^2 s}{4c} \sum_{q,q'=t,b} Q_q v_{q'} \Pi_{qq,q'q'}^{V}(q^2),
\end{align}
with $v_q=2 I_3^q - 4 s^2 Q_q$, where $I_3^q, Q_q$ are the weak
isospin and the electric charge, respectively.
$\Pi_{q_1q_2,q_3q_4}^V, \Pi_{q_1q_2,q_3q_4}^A$ denote the transverse
vector and axialvector polarization functions of correlators of one
current with quark flavours $q_1, q_2$ and a second current with
flavours $q_3, q_4$. It is customary to
single out (pure) flavour singlet contributions, where the external
bosons couple to different quark lines. For the non-singlet
contributions we obtain $\Pi_{qb,qb}^V(q^2) = \Pi_{qb,qb}^A(q^2)$
neglecting the bottom-quark mass.

At the three-loop level, many terms in the desired low-energy
expansions around $q^2 = 0$ are known for all correlators whose
currents involve at least one massive
quark~\cite{Chetyrkin:1997qi,Chetyrkin:1997mb,Chetyrkin:2000mq,Chetyrkin:2001je,Boughezal:2006uu,Maier:2007yn,Maier:2011jd}. These
have been used to obtain high-precision predictions for arbitrary
kinematics for the diagonal non-singlet
correlators~\cite{Chetyrkin:1996cf,Hoang:2008qy,Kiyo:2009gb,Greynat:2010kx,Maier:2017ypu},
and for the diagonal vector correlator even the analytic result is
known~\cite{Chen:2024amk}. At four-loop order, most required
ingredients are available as well. For the diagonal axialvector and
vector non-singlet polarization functions five terms in the low-energy
expansions have been
calculated~\cite{Chetyrkin:2006xg,Boughezal:2006px,Maier:2008he,Sturm:2008eb,Maier:2009fz,Kiyo:2009gb,Maier:2017ypu},
and approximations for arbitrary energies have been derived
in~\cite{Hoang:2008qy,Kiyo:2009gb,Greynat:2010kx,Greynat:2011zp}. Similarly,
six terms are known for the low-energy expansion of the non-diagonal
polarization function $\Pi^V_{tb,tb}$
\cite{Maier:2015jxa}. $\Pi^V_{bb,bb}$ has been given
in~\cite{Chetyrkin:2010dx} in the limit of an infinite top-quark mass.

Hence, only two building blocks are missing to obtain the four-loop
QCD corrections to $\Delta \bar{r}$ and $\Delta \kappa$ beyond the limit of
an infinitely heavy top quark:
\begin{enumerate}
\item The four-loop corrections to the flavour-singlet polarization
functions. Note that for the vector polarization function the first
three expansion terms are already available~\cite{Maier:2008he}.
\item Top-mass corrections to the diagonal bottom-quark correlator
$\Pi^V_{bb,bb} = \Pi^A_{bb,bb}$.
\end{enumerate}
In this work, we provide these two missing ingredients as well as the
sixth term in the low-energy expansion of the axialvector polarization
function. We then discuss the resulting predictions for $\Delta \bar{r}$
and $\Delta \kappa$ at four loops to order $\frac{M_Z^{10}}{m_t^{10}}$.

\section{Calculation and results}
\label{sec:res}

We use a standard setup for the calculation. First, we generate the
Feynman diagrams using \texttt{QGRAF}~\cite{Nogueira:1991ex}. Within
an asymptotic large mass
expansion~\cite{Smirnov:1990rz,Smirnov:1994tg,Tkachov:1994gdc,Smirnov},
we obtain both purely hard contributions and products of hard and soft
integrals. For the former, we identify the underlying vacuum integral
families with up to four loops using custom code~\cite{dynast} based
on \texttt{nauty} and \texttt{Traces}~\cite{mckay2013practical} and
then insert the Feynman rules using FORM~\cite{Davies:2026cci}, in
particular the \texttt{color}~\cite{vanRitbergen:1998pn} package for
the colour algebra. For the remaining contribution to the asymptotic
expansion, we first insert the Feynman rules and factorize the
diagrams before identifying the massive vacuum and massless
propagator-type integral families with at most three loops. We use
Larin's prescription~\cite{Larin:1993tq} to treat $\gamma_5$ inside
singlet diagrams and naively anticommute it in all other
diagrams. After expanding in the external momentum and performing
tensor reduction with \texttt{OPITeR}~\cite{Goode:2024cfy}, we reduce
the resulting scalar integrals to a small set of known basis
integrals~\cite{Laporta:2002pg,Chetyrkin:2004fq,Schroder:2005va,Kniehl:2005yc,Schroder:2005db,Schroder:2005hy,Chetyrkin:2006dh,Bejdakic:2006vg,Kniehl:2006bf,Kniehl:2006bg}
using a custom implementation~\cite{crusher} of integration-by-parts
reduction~\cite{Chetyrkin:1981qh} via Laporta's
algorithm~\cite{Laporta:2000dsw}. Here, we employ finite field
techniques~\cite{Kauers:2008zz,Kant:2013vta,vonManteuffel:2014ixa,Peraro:2016wsq,Klappert:2019emp}
implemented within \texttt{tinbox}~\cite{tinbox}.

Full semi-analytic results for $\Delta \bar{r}$ and $\Delta \kappa$ up
to $\frac{M_Z^{10}}{m_t^{10}}$ for both $\overline{\text{MS}}$ and
on-shell top quark masses are provided in electronic form as
supplementary material to this work. For completeness, we further
include results for the vector and axialvector polarization functions,
as well as for the polarization function of scalar currents and the
longitudinal axialvector polarization function $\Pi^A_L$, from which
the pseudoscalar polarization function $\Pi^P$ can be obtained via the
Ward identity
\begin{equation}
  \label{eq:PiP_WI}
  \Pi^P(q^2) - q^2\frac{\partial \Pi^P(q^2)}{\partial q^2}\bigg\rvert_{q^2=0} = \frac{q^2}{4m^2} \Pi^A_L(q^2).
\end{equation}

In the following, we only show numeric results for $\Delta \bar{r}$
and $\Delta \kappa$ expressed in terms of the $\overline{\text{MS}}$
top-quark mass $\bar{m}_t$ for $\mu = \bar{m}_t$ and omit the
contributions of order $\frac{M_Z^6}{\bar{m}_t^6}$ and higher for the
sake of readability. We obtain
\begin{equation}
  \label{eq:rMS}
  \begin{split}
    & - \frac{8 \sqrt{2} \pi^2 s^2}{3 G_F c^2} \Delta \bar{r}_{\overline{\mathrm{MS}}} =
1.000
+ \frac{\alpha _s}{\pi }\Bigg[-0.1932\Bigg]
+ \frac{\alpha _s^2}{\pi ^2}\Bigg[-3.970\Bigg]
+ \frac{\alpha _s^3}{\pi ^3}\Bigg[-1.680\Bigg]
\\&
+ \frac{M_Z^2}{\bar{m}_t^2}\Bigg\{0.3333-0.5926 s^2
+ \log \left(\frac{M_Z^2}{\bar{m}_t^2}\right)\bigg[0.8889 s^2-0.6667\bigg]
\\&
+ \frac{\alpha _s}{\pi }\Bigg[0.2137-2.268 s^2
+ \log \left(\frac{M_Z^2}{\bar{m}_t^2}\right)\bigg[0.8889 s^2-0.6667\bigg]\Bigg]
\\&
+ \frac{\alpha _s^2}{\pi ^2}\Bigg[7.204 s^2-5.649
+ \log \left(\frac{M_Z^2}{\bar{m}_t^2}\right)\bigg[1.253 s^2+0.08829\bigg]
+ \log ^2\left(\frac{M_Z^2}{\bar{m}_t^2}\right)\bigg[0.4722-0.8519 s^2\bigg]\Bigg]
\\&
+ \frac{\alpha _s^3}{\pi ^3}\Bigg[38.09 s^2-14.14
+ \log \left(\frac{M_Z^2}{\bar{m}_t^2}\right)\bigg[30.95-38.57 s^2\bigg]
\\&
+ \log ^2\left(\frac{M_Z^2}{\bar{m}_t^2}\right)\bigg[0.01596-3.475 s^2\bigg]
+ \log ^3\left(\frac{M_Z^2}{\bar{m}_t^2}\right)\bigg[1.088 s^2-0.6034\bigg]\Bigg]\Bigg\}
\\&
+ \frac{M_Z^4}{\bar{m}_t^4}\Bigg\{-1.000 s^4+1.144 s^2-0.4000
+ \frac{\alpha _s}{\pi }\Bigg[-0.9155 s^4+0.5216 s^2-0.2452\Bigg]
\\&
+ \frac{\alpha _s^2}{\pi ^2}\Bigg[-0.3878 s^4-0.7808 s^2+0.1760
+ \log \left(\frac{M_Z^2}{\bar{m}_t^2}\right)\bigg[0.05794 s^2-0.07226\bigg]
\\&
+ \log ^2\left(\frac{M_Z^2}{\bar{m}_t^2}\right)\bigg[0.004938-0.006584 s^2\bigg]\Bigg]
+ \frac{\alpha _s^3}{\pi ^3}\Bigg[10.47 s^4-13.40 s^2+4.591
\\&
+ \log \left(\frac{M_Z^2}{\bar{m}_t^2}\right)\bigg[0.3775-0.6722 s^2\bigg]
+ \log ^2\left(\frac{M_Z^2}{\bar{m}_t^2}\right)\bigg[0.02813 s^2-0.04243\bigg]
\\&
+ \log ^3\left(\frac{M_Z^2}{\bar{m}_t^2}\right)\bigg[0.003475-0.002987 s^2\bigg]\Bigg]\Bigg\}
\end{split}
\end{equation}
\begin{equation}
  \label{eq:kMS}
  \begin{split}
 & \frac{8 \sqrt{2} \pi^2 s^2}{3 G_F c^2}   \Delta \kappa_{\overline{\text{MS}}} =
1.000
+ \frac{\alpha _s}{\pi }\Bigg[-0.1932\Bigg]
+ \frac{\alpha _s^2}{\pi ^2}\Bigg[-3.970\Bigg]
+ \frac{\alpha _s^3}{\pi ^3}\Bigg[-1.680\Bigg]
\\&
+ \frac{M_Z^2}{\bar{m}_t^2}\Bigg\{0.3333-0.2963 s^2
+ \log \left(\frac{M_Z^2}{\bar{m}_t^2}\right)\bigg[0.4444 s^2-0.6667\bigg]
+ \frac{\alpha _s}{\pi }\Bigg[0.2137-1.134 s^2
\\&
+ \log \left(\frac{M_Z^2}{\bar{m}_t^2}\right)\bigg[0.4444 s^2-0.6667\bigg]\Bigg]
+ \frac{\alpha _s^2}{\pi ^2}\Bigg[3.602 s^2-5.649
\\&
+ \log \left(\frac{M_Z^2}{\bar{m}_t^2}\right)\bigg[0.6263 s^2+0.08829\bigg]
+ \log ^2\left(\frac{M_Z^2}{\bar{m}_t^2}\right)\bigg[0.4722-0.4259 s^2\bigg]\Bigg]
\\&
+ \frac{\alpha _s^3}{\pi ^3}\Bigg[19.04 s^2-14.14
+ \log \left(\frac{M_Z^2}{\bar{m}_t^2}\right)\bigg[30.95-19.29 s^2\bigg]
+ \log ^2\left(\frac{M_Z^2}{\bar{m}_t^2}\right)\bigg[0.01596-1.737 s^2\bigg]
\\&
+ \log ^3\left(\frac{M_Z^2}{\bar{m}_t^2}\right)\bigg[0.5442 s^2-0.6034\bigg]\Bigg]\Bigg\}
+ \frac{M_Z^4}{\bar{m}_t^4}\Bigg\{-0.5000 s^4+0.8222 s^2-0.4000
\\&
+ \frac{\alpha _s}{\pi }\Bigg[-0.4577 s^4+0.4897 s^2-0.2452\Bigg]
+ \frac{\alpha _s^2}{\pi ^2}\Bigg[-0.1939 s^4-0.2935 s^2+0.1760
\\&
+ \log \left(\frac{M_Z^2}{\bar{m}_t^2}\right)\bigg[0.02897 s^2-0.07226\bigg]
+ \log ^2\left(\frac{M_Z^2}{\bar{m}_t^2}\right)\bigg[0.004938-0.003292 s^2\bigg]\Bigg]
\\&
+ \frac{\alpha _s^3}{\pi ^3}\Bigg[5.234 s^4-9.316 s^2+4.591
+ \log \left(\frac{M_Z^2}{\bar{m}_t^2}\right)\bigg[0.3775-0.3361 s^2\bigg]
\\&
+ \log ^2\left(\frac{M_Z^2}{\bar{m}_t^2}\right)\bigg[0.01406 s^2-0.04243\bigg]
+ \log ^3\left(\frac{M_Z^2}{\bar{m}_t^2}\right)\bigg[0.003475-0.001494 s^2\bigg]\Bigg]\Bigg\}
\\&
    + i \Bigg (
+ \frac{M_Z^2}{\bar{m}_t^2}\Bigg\{1.396 s^2-1.862 s^4
+ \frac{\alpha _s}{\pi }\Bigg[1.396 s^2-1.862 s^4\Bigg]
+ \frac{\alpha _s^2}{\pi ^2}\Bigg[1.968 s^2-2.624 s^4
\\&
+ \log \left(\frac{M_Z^2}{\bar{m}_t^2}\right)\bigg[3.568 s^4-2.676 s^2\bigg]\Bigg]
+ \frac{\alpha _s^3}{\pi ^3}\Bigg[35.79 s^4-26.84 s^2
\\&
+ \log \left(\frac{M_Z^2}{\bar{m}_t^2}\right)\bigg[14.56 s^4-10.92 s^2\bigg]
+ \log ^2\left(\frac{M_Z^2}{\bar{m}_t^2}\right)\bigg[5.129 s^2-6.839 s^4\bigg]\Bigg]\Bigg\}
\\&
+ \frac{M_Z^4}{\bar{m}_t^4}\Bigg\{
+ \frac{\alpha _s^2}{\pi ^2}\Bigg[0.09102 s^2-0.1214 s^4
+ \log \left(\frac{M_Z^2}{\bar{m}_t^2}\right)\bigg[0.02758 s^4-0.02069 s^2\bigg]\Bigg]
\\&
+ \frac{\alpha _s^3}{\pi ^3}\Bigg[1.934 s^4-1.148 s^2
+ \log \left(\frac{M_Z^2}{\bar{m}_t^2}\right)\bigg[0.08836 s^2-0.1178 s^4\bigg]
\\&
+ \log ^2\left(\frac{M_Z^2}{\bar{m}_t^2}\right)\bigg[0.01877 s^4-0.01408 s^2\bigg]\Bigg]\Bigg\}
    \Bigg)
  \end{split}
\end{equation}
\section{Numerical impact}
\label{sec:numerics}

We now assess the numerical impact of the various coefficients in the
perturbative and the $M_W \sim M_Z \ll m_t$ expansion on the W boson
mass and the weak mixing angle for $\mu = m_t$. To this end, we solve
equation~\eqref{eq:deltar_def} numerically with $G_F = 1.1663785
\times 10^{-5}, M_Z = 91.1879\,\text{GeV}, \alpha = 1/137.035999177,
\alpha_s^{(n_f = 5)}(M_Z) = 0.1180$~\cite{PDG}. We neglect the
difference between $\Delta r$ and $\Delta \bar{r}$,
c.f.~equation~(\ref{eq:deltarbar}), and take the difference between
the W mass obtained when expanding $\Delta \bar{r}$ up to some given
powers $n, m$ in $\alpha_s$ and $M_Z^2/m_t^2$ and the resulting mass
when removing the term proportional to $\alpha_s^n
(M_Z^2/m_t^2)^m$. In table~\ref{tab:delta_MW_OS} we show the results
for the shift in the W boson mass for an on-shell top quark mass of
$M_t = 172.6\,$GeV~\cite{PDG} and $\alpha_s(M_t) = 0.1077$, which we
obtain via five-loop running
\cite{Baikov:2016tgj,Herzog:2017ohr,Luthe:2017ttg,Chetyrkin:2017bjc}
and four-loop on-shell
decoupling~\cite{Schroder:2005hy,Chetyrkin:2005ia} at $\mu =
M_t$. Table~\ref{tab:delta_MW_MS} shows the corresponding values for
an $\MS$ top quark mass of $\bar{m}_t(\bar{m}_t) =
162.5\,$GeV~\cite{PDG} and $\alpha_s(\bar{m}_t) = 0.1085$.
\begin{table}[h]
  \centering
  \begin{tabular}{lrrrr}
    \toprule
     $\Delta M_W$\,[MeV] & $\alpha_s^0$ & $\alpha_s^1$ & $\alpha_s^2$ & $\alpha_s^3$ \\
    \midrule
    $M_t^2$              & 539.6253     & -55.0977     & -9.5853      & -2.0953      \\
    $M_t^0$              & 134.0347     & -5.0638      & -2.3777      & -0.7162      \\
    $M_t^{-2}$           & -9.5827      & -1.1332      & -0.2217      & -0.0724      \\
    $M_t^{\le -4}$       & -0.6911      & -0.1172      & -0.0347      & -0.0124      \\
    \bottomrule
  \end{tabular}
  \caption{Change of $M_W$ in MeV due to various orders in the perturbative and large on-shell top mass expansion.}
  \label{tab:delta_MW_OS}
\end{table}

\begin{table}[h]
  \centering
  \begin{tabular}{lrrrr}
    \toprule
     $\Delta M_W$\,[MeV] & $\alpha_s^0$ & $\alpha_s^1$ & $\alpha_s^2$ & $\alpha_s^3$ \\
    \midrule
    $\bar{m}_t^2$        & 475.8288 & -3.3032 & -2.3425 & -0.0342  \\
    $\bar{m}_t^0$        & 122.7257 & 1.7872  & -0.8006 & -0.2218  \\
    $\bar{m}_t^{-2}$     & -10.6056 & -0.3080 & 0.0047  & 0.0041   \\
    $\bar{m}_t^{\le -4}$ & -0.8721  & 0.0182  & -0.0003 & 0.0004  \\
    \bottomrule
  \end{tabular}
  \caption{Change of $M_W$ in MeV due to various orders in the perturbative and large $\MS$ top mass expansion.}
  \label{tab:delta_MW_MS}
\end{table}

Comparing to the expected precision at the FCC-ee, we observe that the
size of the new corrections of order $\alpha_s^3 m_t^{\leq 0}$ in
either scheme clearly exceed both the projected systematic uncertainty
of $\Delta M_W = 0.16\,$MeV and the statistical uncertainty of
$\Delta M_W = 0.18\,$MeV~\cite{FCC:2025lpp}. Assuming that the
apparent convergence pattern continues to higher orders, we estimate
that the order $\alpha_s^4 M_t^2$ mass shift will amount to several
hundred keV. This provides a strong motivation to compute the
five-loop QCD corrections to the $\rho$ parameter.

We further note that the mass shifts in the \MS scheme from order
$\alpha_s^1$ on are substantially smaller than in the on-shell
scheme. However, we also find that the newly computed $\alpha_s^3
\bar{m}_t^0$ contribution exceeds the formally leading term of order
$\alpha_s^3 \bar{m}_t^2$.

To analyse the numerical impact of the corrections to the weak mixing
angle we show the respective contributions to $\Re(\Delta \kappa)$,
i.e.\ the relative change in the real part of $\sin^2 \theta_w$
compared to the baseline $\sin^2 \theta_w = 1 -
\frac{M_W^2}{M_Z^2}$. In addition to the above parameters we use $M_W
= 80.3625\,$GeV~\cite{PDG}. We obtain table~\ref{tab:delta_s_OS} for
the on-shell and table~\ref{tab:delta_s_MS} for the $\MS$ top quark
mass. The general pattern is similar to $\Delta \bar{r}$,
c.f.~tables~\ref{tab:delta_MW_OS} and \ref{tab:delta_MW_MS}, with
smaller corrections in the \MS scheme apart from a rather big
contribution at order $\alpha_s^3 \bar{m}_t^0$.

\begin{table}[h]
  \centering
  \begin{tabular}{lrrrr}
    \toprule
     $\Re(\Delta \kappa)\,[\permil]$ & $\alpha_s^0$ & $\alpha_s^1$ & $\alpha_s^2$ & $\alpha_s^3$ \\
    \midrule
    $M_t^2$                          & 32.46451     & -3.18297     & -0.55684     & -0.12184     \\
    $M_t^0$                          & 8.98207      & -0.25739     & -0.15307     & -0.04734     \\
    $M_t^{-2}$                       & -0.61033     & -0.06955     & -0.01378     & -0.00457     \\
    $M_t^{\le -4}$                   & -0.04466     & -0.00748     & -0.00223     & -0.00080     \\
    \bottomrule
  \end{tabular}
  \caption{Relative change in $\sin^2 \theta_w$ due to various orders in the perturbative and large on-shell top mass expansion.}
  \label{tab:delta_s_OS}
\end{table}

\begin{table}[h]
  \centering
  \begin{tabular}{lrrrr}
    \toprule
     $\Re(\Delta \kappa)\,[\permil]$ & $\alpha_s^0$ & $\alpha_s^1$ & $\alpha_s^2$ & $\alpha_s^3$ \\
    \midrule
    $\bar{m}_t^2$                    & 28.77623     & -0.19207     & -0.13628     & -0.00199     \\
    $\bar{m}_t^0$                    & 8.36201      & 0.19283      & -0.04978     & -0.01509     \\
    $\bar{m}_t^{-2}$                 & -0.68856     & -0.01563     & 0.00062      & 0.00028      \\
    $\bar{m}_t^{\le -4}$             & -0.05763     & 0.00150      & 0.00001      & 0.00003      \\
    \bottomrule
  \end{tabular}
  \caption{Relative change in $\sin^2 \theta_w$ due to various orders in the perturbative and large $\MS$ top mass expansion.}
  \label{tab:delta_s_MS}
\end{table}

\section{Conclusions}
\label{sec:conclusions}

We have computed four-loop QCD corrections involving top quarks to
relations between the weak boson masses, the Fermi and finestructure
constants, and the weak mixing angle. Our results are obtained
in a large top mass expansion with a negligible truncation error. We
find that the corrections are numerically relevant in the light of the
expected experimental uncertainties at the FCC-ee. We estimate that
at least in the on-shell scheme this will also hold true for the so
far unknown five-loop corrections, which makes their computation
highly desirable. Concretely, the shift in the predicted value of the
W boson mass due to our new results amounts to approximately
$-0.8$\,MeV in the on-shell scheme and $-0.2$\,MeV in the \MS scheme,
to be compared to projected experimental statistical and systematic
uncertainties of about $0.2$\,MeV each.

Semi-analytic results for $\Delta \bar{r}, \Delta \kappa,$ and fully
analytic results for the moments of the flavour-diagonal vector,
axialvector, scalar, and pseudoscalar polarization functions up to
four loops are provided as supplementary material to this paper.

\section*{Acknowledgements}

We thank Matthias Steinhauser for initial discussions motivating this
work and for comments on the manuscript.

\bibliographystyle{JHEP}
\bibliography{biblio}

\end{document}